\documentclass[a4paper,11pt]{article}

\usepackage{rotating}
\usepackage{jheppub} 

\usepackage{amssymb}
\usepackage{amsmath}
\usepackage{amsfonts}
\usepackage{graphicx}
\usepackage{color}
\usepackage{xcolor} 
\usepackage{xspace}
\usepackage{ulem}
\usepackage{lscape}
\usepackage{wasysym} 
\usepackage{multirow,rotating}
\usepackage{mathrsfs} 

\def\gsim{\raise0.3ex\hbox{$\;>$\kern-0.75em\raise-1.1ex\hbox{$\sim\;$}}}
\def\lsim{\raise0.3ex\hbox{$\;<$\kern-0.75em\raise-1.1ex\hbox{$\sim\;$}}}

\newcommand{\ba}[1]{\begin{eqnarray} \label{(#1)}}
\newcommand{\ea}{\end{eqnarray}}

\renewcommand{\vec}[1]{\boldsymbol{{\mathrm #1}}}

\usepackage[labelfont=bf]{caption}
\usepackage{tabularx,booktabs}
\usepackage{hhline}
\usepackage{multicol}
\usepackage{array,multirow}
\usepackage{booktabs}
\usepackage{colortbl}
\usepackage{float}
\usepackage{verbatimbox}
\usepackage{adjustbox}

\usepackage{tabularray} 

\allowdisplaybreaks[4]

\title{
\vspace*{2cm}

$n \rightarrow K \ell$ and the baryon asymmetry of the universe}

\author[a]{Carolina Arbel\'aez,}
\author[b,c]{Juan Carlos Helo,}
\author[d]{Martin Hirsch,}
\author[b,c]{Toshihiko Ota}

\emailAdd{carolina.arbelaez@usm.cl}
\emailAdd{jchelo@userena.cl}
\emailAdd{mahirsch@ific.uv.es}
\emailAdd{toshihiko.ota@userena.cl}
  
\affiliation[a]{Universidad T\'ecnica Federico Santa Mar\'ia
 and Centro Cient\'ifico Tecnol\'ogico\\
 de Valpara\'iso CCTVal, Casilla 110-V, Valpara\'iso, Chile}
   
\affiliation[b]{Departamento de F\'{i}sica, Facultad de Ciencias,
  Universidad de La Serena, Avenida Cisternas 1200, La Serena, Chile}

\affiliation[c]{Millennium Institute for Subatomic Physics at the High
  Energy Frontier (SAPHIR), Fern\'{a}ndez Concha 700, Santiago, Chile}

\affiliation[d]{Instituto de F\'{\i}sica Corpuscular
  (CSIC-Universitat de València),  46980 Paterna, Spain}


\abstract{%
  The observed baryon asymmetry (BAU) of the universe puts strong constraints on any $(B-L)$-violating interaction. An observation of a $(B-L)$-violating nucleon decay channel would therefore have profound implications for our understanding of the BAU. Here we point out that the observation of the final state with a kaon and a charged lepton in a future nucleon decay experiment would hint at $(B-L)$ violation even if the charge of the lepton is not determined experimentally. In SMEFT, this follows from the fact that \(n\to K^+\ell^-\) arises already at dimension seven, while the $(B-L)$-conserving decay \(n\to K^-\ell^+\) requires dimension-ten operators that, in addition, would be accompanied by lower-dimensional $(B+L)$-violating decay modes. An observation of \(n\to K\ell\) in the absence of other modes such as \(p \to \pi^{0} \ell^+\), would then strongly suggest that $(B-L)$ is violated.}

\makeatletter
\gdef\@fpheader{\phantom{a}}
\makeatother

\begin{document}
\maketitle


\section{Introduction}
\label{Sec:intro}

Proton decay searches often focus on $(B-L)$-conserving final states,
such as $p\to \pi^0 e^+$ or $p \to K^+ \overline{\nu}$, since these
are the modes predicted by minimal $SU(5)$~\cite{Georgi:1974sy} and
supersymmetric versions of $SU(5)$ and $SO(10)$~\cite{Nath:2006ut}. 
However, while much less-known there are also non-SUSY $SO(10)$ models 
that allow for $(B-L)$-violating nucleon decays to dominate, 
as discussed, for example, in \cite{Babu:2012iv}.
More agnostically, from the point of view of Standard Model Effective
Field Theory (SMEFT), cf. e.g.,~\cite{Isidori:2023pyp,Aebischer:2025qhh} for reviews, 
nucleon decays are described by a series of
non-renormalizable operators. On studies of nucleon decays in light of effective operators, cf. e.g.,~\cite{Bowes:1996xy,Kovalenko:2002eh,deGouvea:2014lva,Kobach:2016ami,Assad:2017iib,Hambye:2017qix,Fonseca:2018ehk,Helo:2019yqp,Heeck:2019kgr,Chamoun:2020aft,Dorsner:2022twk,Beneito:2023xbk,Gargalionis:2024nij,Heeck:2024jei,Hamaguchi:2024ewe,Fan:2024gzc,IBeneito:2025nby,Liao:2025vlj,Heeck:2025btc,Liao:2025sqt,Chen:2025mjt,Heeck:2026dmh,Liao:2026ugl,Fan:2026fqo,Heeck:2026ked,Fan:2026csl}.\footnote{%
On the pioneering works in this aspect, cf.~\cite{Weinberg:1979sa,Wilczek:1979hc,Weldon:1980gi,Weinberg:1980bf,Abbott:1980zj,Wise:1980ch,Claudson:1981gh,Rao:1983sd}.
For studies of nucleon decays from the viewpoint of effective operators, cf. e.g.,~\cite{Davoudiasl:2014gfa,Helo:2018bgb,Fajfer:2020tqf,Liang:2023yta,Fridell:2023tpb,Domingo:2024qoj,Li:2024liy,Li:2025slp,Heeck:2025uwh,Fan:2025xhi,Helo:2025kgx,Ma:2025mjy,Adolf:2026khc}.}

It is easy to show within SMEFT that $(B+L)$-violating nucleon decays are
generated by operators with even dimensions ($d=6,8,\cdots$), while
$(B-L)$-violating ones require operators with odd dimensions,
starting at $d=7$~\cite{Kobach:2016ami}. 
Since $(B-L)$ violation requires higher dimensional
operators than $(B+L)$-violating nucleon decays, it seems reasonable to
concentrate on the latter, completely independent of the preferences
of GUT models. 
However, $(B+L)$ and $(B-L)$ violation have very different
implications for the observed baryon asymmetry of the universe (BAU),
and this motivates a more detailed study of $(B-L)$-violating nucleon decays.

The origin of the BAU is still unknown. Sakharov formulated three
conditions for a successful baryogenesis
\cite{Sakharov:1967dj}: (i) baryon number violation; (ii) departure
from thermal equilibrium and (iii) CP-violation.  
In the Standard Model (SM) baryon number is violated non-perturbatively 
by sphaleron interactions \cite{Manton:1983nd,Klinkhamer:1984di}, 
thus fulfilling Sakharov's first condition. 
However, at temperatures above the electroweak scale these interactions 
are in equilibrium \cite{Kuzmin:1985mm} and thus sphalerons by themselves 
cannot explain the BAU.

Sphalerons conserve $(B-L)$ and this has important consequences for any
high-scale model of baryogenesis. 
On the one hand, any model that can generate a net $(B-L)$ number in 
the early universe, can lead to a successful generation of the BAU, 
since sphalerons will convert a $(B-L)$ asymmetry into a $(B+L)$ one.  
The most prominent example for such a scenario is the famous leptogenesis mechanism
\cite{Fukugita:1986hr,Flanz:1994yx,Covi:1996wh,Pilaftsis:1997jf}, but
other models have been discussed in the literature as well 
\cite{Enomoto:2011py,Babu:2012vb,Babu:2012iv,Hati:2018cqp}.
On the other hand, if $(B-L)$-violating interactions are strong enough 
to come into equilibrium at any time before the electroweak phase
transition, any pre-existing, non-zero $(B-L)$ will be erased.\footnote{%
If there is a non-zero asymmetry in $L_{\alpha}-L_{\beta}$, flavour
effects may lead to a successful BAU generation, even if the total
$(B-L)$ is zero, see for example
\cite{Dreiner:1992vm,Abada:2006fw,Abada:2006ea,Nardi:2006fx}.}  
This leads to important constraints on the allowed strength of 
$(B-L)$-violating interactions. Motivated by this argument, in this paper we
study the implications that a hypothetical observation of a $(B-L)$-violating
nucleon decay in the next round of proton decay experiments, such as
Hyper-Kamiokande \cite{Hyper-Kamiokande:2018ofw}, DUNE
\cite{DUNE:2020lwj} or JUNO \cite{JUNO:2015zny}, would have for
baryogenesis.

Determining whether $(B-L)$ is violated or not in a nucleon decay
requires measuring the charge of the outgoing lepton (or meson). 
Since none of the existing and forthcoming nucleon decay
experiments is equipped with a magnetic field, this is likely
impossible in practice.\footnote{JUNO \cite{JUNO:2015zny} is designed
to measure neutrino oscillation parameters with high precision using
reactor anti-neutrinos with typical energies of a few MeV. 
The positrons from the inverse beta decay reaction are stopped in the
scintillator and pair-annihilate to give two 511 keV $\gamma$ rays. 
This signal can in principle be used to distinguish electrons
and positrons.  
However, for electrons/positrons with much larger
energies, as expected in nucleon decays, this signal is probably too
weak to be useful. We thank Michael Wurm for an enlightening discussion
on this subject.}  
Consequently, proton decay experiments have paid little attention to 
$(B-L)$-violating final states.\footnote{%
Super-Kamiokande has derived limits on, for example,
$K^+ \overline{\nu}$ \cite{Super-Kamiokande:2025ibz}, which equally
apply to the $(B-L)$-violating final state $K^+ \nu$. But there
are no dedicated searches for negatively charged leptons in nucleon
decay by Super-Kamiokande.}

Here, we point out that an observation of the decay $n \to K \ell$, 
where $\ell \in \{e,\mu\}$, 
would favor $(B-L)$ violation, even if the charge of the lepton
(and/or kaon) is not determined experimentally.  
Our argument is based on an operator analysis in SMEFT. 
As we show below, in SMEFT the mode $n \to K^+ \ell^-$ does occur at
$d=7$, whereas $n \to K^- \ell^+$ requires at least a $d=10$ operator. 
Consequently,
the $(B-L)$-violating mode is sensitive to new physics scales more than
5 orders of magnitude larger than the $(B+L)$-violating one. 
Moreover, we argue that there is no symmetry that allows the $d=10$ operator,
while simultaneously forbidding the $d=6$ and $d=8$ $(B+L)$-violating
operators.  
Thus, $n \to K^- \ell^+$ should always be accompanied by
other decay modes, such as $p \to \pi^0 e^+$, which cannot be
generated by $(B-L)$-violating operators. 
Non-observation of these final states would strengthen the interpretation 
of the $(B-L)$ origin, if $n \to K \ell$ were to be observed. 
We admit, however, that our argument is mainly theoretical and not completely 
loop-hole free. 
In fact, a half-live of order $10^{31}$ yr for $n \to K^- \ell^+$ requires 
a $d=10$ operator with a scale of roughly $\sim 110$ TeV, which is not excluded
by current experimental data, of course. 
We will discuss the above arguments in much more detail in Section~\ref{sect:EFT} 
of this paper.

We also emphasize that Super-Kamiokande has not searched for neutron decays
into final states with charged kaons. 
Thus, the currently best limits on $n \to K^- \ell^+$ (from IMB-3~\cite{McGrew:1999nd}) 
and $n \to K^+ \ell^-$ (from Frejus~\cite{Frejus:1991ben}) are rather old and
only of order ${\cal O}(10^{31})$ yr. 
A dedicated search by Super-Kamiokande should be able to improve these limits 
(or discover nucleon decay!) by roughly two orders of magnitude. 
Even better prospects exist for Hyper-Kamiokande~\cite{Hyper-Kamiokande:2018ofw}, 
DUNE~\cite{DUNE:2020lwj} and JUNO~\cite{JUNO:2015zny}. 
Hyper-Kamiokande's huge effective volume will generated unprecedented statistics,
while both DUNE and JUNO should be particularly sensitive to nucleon
decay modes involving kaons, due to their lower energy thresholds.

We discuss the implications of a discovery of $n\to K\ell$ for high-scale 
baryogenesis scenarios in more detail in Section 3. 
If the effective operator that is responsible for $n \rightarrow K^{+} \ell^{-}$ came into 
thermal equilibrium in early universe, it washed out any pre-existing $(B-L)$ number.
Together with the wash out of $(B+L)$ number by the sphaleron processes, 
we can conclude that any baryon number generated at high temperature 
could not survive until today.
In short, the discovery of $n \rightarrow K \ell$ would point to baryogenesis at 
intermediate or low scales, where the relevant $(B-L)$-violating interactions remain 
out of equilibrium before the electroweak phase transition ~\cite{Kuzmin:1985mm,Cohen:1990py,Cohen:1990it}.


The rest of this paper is organized as follows. 
In Section 2 we discuss the SMEFT operator analysis for $d = 6$ and $d = 7$ 
operators, leading to nucleon decay. 
We show that $n \to K^-\ell^+$ is not generated at $d = 6$, 
whereas $n \to K^+\ell^-$ appears at $d = 7$, and that one must go up 
to $d = 10$ operators to generate $n \to K^-\ell^+$. 
We demonstrate that $d = 6$ and $d = 8$ operators should coexist with 
these $d = 10$ operators and thus $n\to K^-\ell^+$ must be accompanied 
by other nucleon decay modes, such as $p \to \pi^0e^+$. 
In Section 3 we then draw a general conclusion on baryogenesis scenarios
from a possible discovery of $n \rightarrow K \ell$
in the next round of experiments.
We then close with a brief discussion.

\section{$n\to K \ell$ in SMEFT\label{sect:EFT}}

In this section we study the nucleon decay modes $n\to K^+ \ell^-$ and $n\to K^- \ell^+$ 
in SMEFT, first considering contributions up to  $d=7$. 
Since there is no operator at $d=6$, which generates $n\to K^- \ell^+$, 
we extend the search to higher even dimensions and find that at least a $d=10$ 
operator is required to generate the decay $n\to K^-\ell^+$. 
The implications of this result are discussed below.

SMEFT operators have been listed systematically up to $d=12$ in the
literature. 
One can find the list of $d=6$ operators in \cite{Buchmuller:1985jz,Grzadkowski:2010es}, 
$d=7$ in \cite{Lehman:2014jma,Liao:2016hru}, 
$d=8$ in \cite{Murphy:2020rsh,Li:2020gnx}, 
$d=9$ in \cite{Li:2020xlh,Liao:2020jmn}, and all operators up to 
$d=12$ in \cite{Harlander:2023psl}.\footnote{%
Tree-level realizations of SMEFT operators can be found up to $d=8$
in \cite{deBlas:2017xtg,Li:2022abx,Li:2023cwy,Li:2023pfw}.
As for baryon-number-violating operators that do not contain a derivative,
the realizations are listed up to $d=15$~\cite{Heeck:2026dmh}.}
It is known that a simple relation, which is important for the present work, 
exists between the dimension $d$ of any SMEFT operator and 
its baryon and lepton number. 
The relation was proven by Kobach in \cite{Kobach:2016ami}, 
and can be summarized as
\begin{align}
 \text{$d$ is even}&\longleftrightarrow |\Delta B - \Delta L|=0,4,8,12, \cdots,
 \\
 \text{$d$ is odd}&\longleftrightarrow |\Delta B - \Delta L|=2,6,10,14, \cdots.
\end{align}
In short, $(B-L)$-conserving processes are related to effective
operators with $d=6, 8, 10, \cdots$. 
On the other hand, the process $n \rightarrow K^{+} \ell^{-}$, which is 
$|\Delta B - \Delta L| = 2$, can only be generated via effective operators 
with $d=7,9,11,\cdots$. 
We thus first discuss SMEFT at $d=6$ and $d=7$. 

\subsection{SMEFT at $d\le 7$\label{subsect:d7}}

In Tab.~\ref{Tab:d6d7channels} we list all baryon number violating
$d=6$ and $d=7$ SMEFT operators, together with the relevant two-body
nucleon decay processes. We indicate by a check-mark, which decay
processes a given operator can generate. 
We have suppressed generation indices of the operators for simplicity. 
At $d=6$ all operators automatically generate more than one final state; 
this is no longer the case at $d=7$, where two operators induce {\em only}
$n \to K^+ \ell^-$.
For our purposes, the most important point from Tab.~\ref{Tab:d6d7channels} is 
that no $d=6$ operator generates the final state $n\to K^- \ell^+$. 
Thus, observing $n\to K \ell$ implies the presence of an operator with $d\ge 7$.
In Tab.~\ref{Tab:d6d7channels} we also list the current best limits (at 90\% CL)
for the different final states, together with their references.  
Most limits come from the Super-Kamiokande experiment, but there are a few exceptions. 
For $n\to\pi^+\ell^-$, the Particle Data Group (PDG)~\cite{ParticleDataGroup:2024cfk} 
still quotes the IMB result~\cite{Seidel:1988ut}, whereas Super-Kamiokande set 
a stringent limit for the charge-conjugate channel
$n\to\pi^-\ell^+$~\cite{Super-Kamiokande:2017gev}.
In principle, the bound from Super-Kamiokande should also be applicable to $n \rightarrow \pi^{+} \ell^{-}$, since the sign of the electric charge can not be measured using Cherenkov light.
However, Super-Kamiokande has not published any search on the $n\to K \ell$ modes.  
Therefore the best existing limits on these channels are already more than 
30 years old~\cite{McGrew:1999nd,Seidel:1988ut}. 
A dedicated search by Super-Kamiokande should be able to improve these numbers 
by nearly two orders of magnitude.
\begin{table}[t]
\begin{adjustbox}{width=\columnwidth,center}
\begin{tblr}{ colspec = {c|cccc|cccc}, hline{16} = {dashed} }
\hline
 & \hspace*{-0.5cm}$p \rightarrow$  
     & & & & \hspace*{-0.5cm}$n \rightarrow$ 
		 \\
 $d=6$ 
 & \hspace{0.2cm} 
     $\pi^{0} \ell^{+}$
     \hspace{0.2cm} 
     & \hspace{0.2cm}
	 $K^{0} \ell^{+}$
	 \hspace{0.2cm} 
	 & \hspace{0.2cm}
	     $\pi^{+} \bar{\nu}$
	     \hspace{0.2cm} 
	     & \hspace{0.2cm} $K^{+} \bar{\nu}$
		 \hspace{0.2cm} 
		 & \hspace{0.2cm} 
		     $\pi^{0} \bar{\nu}$
		     \hspace{0.2cm} 
		     & \hspace{0.2cm} 
 $K^{0} \bar{\nu}$ \hspace{0.2cm} 
 & \hspace{0.2cm} $\pi^{-} \ell^{+}$ \hspace{0.2cm} 
 & \hspace{0.2cm} $K^{-} \ell^{+}$ \hspace{0.2cm} 
		 \\
 \SetCell[r=2]{c} $\frac{\text{lifetime}}{10^{30}\text{[yr]}}$ 
 & $24000(e)$ \cite{Super-Kamiokande:2020wjk}
     & $1000(e)$ \cite{Super-Kamiokande:2022egr}
	 & \SetCell[r=2]{c} 390 \cite{Super-Kamiokande:2013rwg}
	     & \SetCell[r=2]{c} 6600 \cite{Mine:2016mxy}
		 & \SetCell[r=2]{c} 1400 \cite{Super-Kamiokande:2025lxa}
		     & \SetCell[r=2]{c} 780 \cite{Super-Kamiokande:2025ibz}
 & $5300(e)$ \cite{Super-Kamiokande:2017gev}
 & $17(e)$ \cite{McGrew:1999nd}
     \\
  & $16000(\mu)$ \cite{Super-Kamiokande:2020wjk}
     & $3600(\mu)$ \cite{Super-Kamiokande:2022egr}
	 &
	     &
		 &
		     &
 & $3500(\mu)$ \cite{Super-Kamiokande:2017gev}
 & $26(\mu)$ \cite{McGrew:1999nd}
     \\
 \hline
$\mathcal{O}_{QQQL}$
& $\checkmark$ & $\checkmark$ & $\checkmark$ & $\checkmark$
& $\checkmark$ & $\checkmark$ & $\checkmark$ &
 \\
$\mathcal{O}_{QQue}$ 
& $\checkmark$ & $\checkmark$ &              &
&              &              & $\checkmark$ &
 \\
$\mathcal{O}_{duue}$
& $\checkmark$ & $\checkmark$ &              &
&              &              & $\checkmark$ &
\\
$\mathcal{O}_{duQL}$
& $\checkmark$ & $\checkmark$ & $\checkmark$ & $\checkmark$
& $\checkmark$ & $\checkmark$ & $\checkmark$ &
 \\
\hline \hline
 $d=7$
 & --- & ---
	 & $\pi^{+} \nu$ & $K^{+} \nu$
		 & $\pi^{0} \nu$
		     & $K^{0} \nu$
 & $\pi^{+} \ell^{-}$ & $K^{+} \ell^{-}$
		 \\
 \SetCell[r=2]{c} 
 $\frac{\text{lifetime}}{10^{30}\text{[yr]}}$ 
 & & & \SetCell[r=2]{c} 390 \cite{Super-Kamiokande:2013rwg}
 & \SetCell[r=2]{c} 6600 \cite{Mine:2016mxy}
 & \SetCell[r=2]{c} 1400 \cite{Super-Kamiokande:2025lxa}
 & \SetCell[r=2]{c} 780 \cite{Super-Kamiokande:2025ibz}
 & $65(e)$ \cite{Seidel:1988ut} & $32(e)$ \cite{Frejus:1991ben}
 \\
 & & & &
 & & & $49(\mu)$ \cite{Seidel:1988ut} & $57(\mu)$ \cite{Frejus:1991ben}
 \\
 \hline
$\mathcal{O}_{\overline{L}dddH}$
&  &  &  & 
&  &  &  & $\checkmark$
 \\
$\mathcal{O}_{\overline{L}dQQ\widetilde{H}}$
&              &              & $\checkmark$ & $\checkmark$ 
& $\checkmark$ & $\checkmark$ &              & $\checkmark$
 \\
$\mathcal{O}_{\overline{e}Qdd\widetilde{H}}$
&  &  &  & 
&  &  &  & $\checkmark$
\\ 
$\mathcal{O}_{\overline{L}dud\widetilde{H}}$
&              &              & $\checkmark$ & $\checkmark$ 
& $\checkmark$ & $\checkmark$ &              & 
\\
$\mathcal{O}_{\overline{L}QdDd}$ 
& & & $\checkmark$ & $\checkmark$
& $\checkmark$ & $\checkmark$ & $\checkmark$ & $\checkmark$
\\
$\mathcal{O}_{\overline{e}dddD}$ 
& & & &
& & & $\checkmark$ & $\checkmark$
\\
\hline
  \end{tblr}
\end{adjustbox}
\caption{List of two-body nucleon decay modes induced by $d=6$ and $d=7$
 SMEFT operators, together with their experimental limits at 90\% CL. 
 Most limits come from the Super-Kamiokande experiment. However,  for
  $n\to \pi^+ \ell^-$ and the $n\to K \ell$ modes, 
 the PDG~\cite{ParticleDataGroup:2024cfk} still cites 
 the results of the IMB~\cite{Seidel:1988ut,McGrew:1999nd} and Frejus~\cite{Frejus:1991ben} 
 experiments.}
\label{Tab:d6d7channels}
\end{table}

Let us estimate the decay rate of $n \rightarrow K^{+} \ell^{-}$
induced by one of the $d=7$ operators.  Here we take
$\mathcal{O}_{\overline{L}dddH}$ as an example
\begin{align}
 \mathscr{L}_{d=7}
 =&
 \frac{c_{d=7}}{\Lambda^{3}}
 \epsilon^{IJK}
 (\overline{L})^{i}_{\dot{a}}
 (d_{R})^{\dot{a}}_{I}
 (\overline{{d_{R}}^{c}})_{J \dot{b}}
 (s_{R})_{K}^{\dot{b}}
 H_{i}
 \nonumber
 \\
 \rightarrow &
 \frac{c_{d=7} \langle H^{0} \rangle}{\Lambda^{3}}
 \epsilon^{IJK}
 (\overline{\ell_{L}})_{\dot{a}}
 (d_{R})^{\dot{a}}_{I}
 (\overline{{d_{R}}^{c}})_{J \dot{b}}
 (s_{R})_{K}^{\dot{b}}
 +
 \cdots
 \label{eq:L-d7-LbardddH}
\end{align}
where $I,J,K$ are the indices for the fundamental representation of
the $SU(3)$ colour (lower (upper) position for $\vec{3}$ ($\overline{\vec{3}}$)), $i$ for $\vec{2}$ of $SU(2)_{L}$, $a,b,...$ for
left-handed 2 spinors, and $\dot{a}, \dot{b}, ...$ for right-handed
2-spinors.  
Note that in our convention the Wilson coefficient
$c_{d=7}$ is dimensionless, and $\Lambda$ is the scale of new physics.
The decay rates of $n \rightarrow K^{+}\ell^{-}$ induced by the
operator Eq.~\eqref{eq:L-d7-LbardddH} can be estimated as
\begin{align}
\Gamma(n \rightarrow K^{+} \ell^{-})
=&
 \frac{|c_{d=7}|^{2} \langle H^{0} \rangle^{2}  |W_{0}|^{2}}{32 \pi \Lambda^{6}}
 m_{n}
 \left(
 1 - \frac{m_{K^{\pm}}^{2}}{m_{n}^{2}}
 \right)^{2}
 \nonumber
 \\
 =&
 \frac{1}
 {3.2 \cdot 10^{31}\text{yr}}
 \left(\frac{c_{d=7}}{1.0} \right)^{2}
 \left(
 \frac{3.0 \cdot 10^{10} \text{GeV}}{\Lambda}
 \right)^{6}
\label{eq:DecayRate-d7-n-to-Kl}
\end{align}
in the limit of a massless $\ell$.  
Here $W_{0}$ is the relevant hadron matrix element, $|W_{0}| = 0.057$ GeV$^{2}$, 
which is given in \cite{Aoki:2017puj}. 
The experimental bounds of $\tau(n\to K^+\ell^-)$ fixes the
combination $|c_{d=7}|/\Lambda^3$. For an order-one Wilson coefficient, 
the current sensitivity corresponds to a new-physics scale of order
$3\cdot 10^{10}\,\mathrm{GeV}$, while a lower scale would require
a suppressed coefficient. For a more precise treatment for the calculation
of the decay rate including RGE effects, matching to the low-energy effective theory (LEFT) 
and the chiral perturbation theory of hadrons, cf.~\cite{Beneito:2023xbk,Liao:2026ugl}.

Finally, we would like to mention that the absence of 
$n \rightarrow K^{-} \ell^{+}$ at $d=6$ was already noted 
by Weinberg in his pioneering paper on the effective field
theory approach of the baryon-number-violating processes~\cite{Weinberg:1979sa}.
In a later paper~\cite{Weinberg:1980bf}, 
he also pointed out that non-derivative $d=7$ operators do not induce 
$n \rightarrow \pi^{+} \ell^{-}$, and instead, 
the final state meson must be a kaon,
if the operators induce a neutron decay with a charged meson and a charged lepton.

\subsection{SMEFT at $d=8$ and $d=10$\label{subsect:d10}}

\begin{table}[t]
\begin{adjustbox}{width=\columnwidth,center}
\begin{tblr}{ colspec = {c|cccc|cccc}, hline{16} = {dashed} }
\hline
 & \hspace*{-0.5cm}$p \rightarrow$  
     & & & & \hspace*{-0.5cm}$n \rightarrow$ 
		 \\
 $d=8$ 
 & \hspace{0.2cm} 
     $\pi^{0} \ell^{+}$
     \hspace{0.2cm} 
     & \hspace{0.2cm}
	 $K^{0} \ell^{+}$
	 \hspace{0.2cm} 
	 & \hspace{0.2cm}
	     $\pi^{+} \bar{\nu}$
	     \hspace{0.2cm} 
	     & \hspace{0.2cm} $K^{+} \bar{\nu}$
		 \hspace{0.2cm} 
		 & \hspace{0.2cm} 
		     $\pi^{0} \bar{\nu}$
		     \hspace{0.2cm} 
		     & \hspace{0.2cm} 
 $K^{0} \bar{\nu}$ \hspace{0.2cm} 
 & \hspace{0.2cm} $\pi^{-} \ell^{+}$ \hspace{0.2cm} 
 & \hspace{0.2cm} $K^{-} \ell^{+}$ \hspace{0.2cm} 
		 \\
 \hline
$\mathcal{O}_{DuddLH}$
& &  & $\checkmark$ & $\checkmark$
& $\checkmark$ & $\checkmark$ &  &
 \\
$\mathcal{O}_{DudQeH}$
& $\checkmark$ & $\checkmark$ &              &
&              &              & $\checkmark$ &
\\
$\mathcal{O}_{DuudLH^{\dagger}}$
& $\checkmark$ & $\checkmark$ &              &
&              &              & $\checkmark$ &
\\
$\mathcal{O}_{DuuQeH^{\dagger}}$
& $\checkmark$ & $\checkmark$ &              &
&              &              & $\checkmark$ &
\\
$\mathcal{O}_{DuQQLH^{\dagger}}$
& $\checkmark$ & $\checkmark$ & $\checkmark$ & $\checkmark$
& $\checkmark$ & $\checkmark$ & $\checkmark$ &
 \\
$\mathcal{O}_{DdQQLH}$
& $\checkmark$ & $\checkmark$ & $\checkmark$ & $\checkmark$
& $\checkmark$ & $\checkmark$ & $\checkmark$ &
 \\
$\mathcal{O}_{DQQQeH}$
& $\checkmark$ & $\checkmark$ &              &
&              &              & $\checkmark$ &
\\
$\mathcal{O}_{uuQLH^{\dagger}H^{\dagger}}$
& $\checkmark$ & $\checkmark$ &              &
&              &              & $\checkmark$ &
\\
$\mathcal{O}_{ddQLHH}$
& &  & $\checkmark$ & $\checkmark$
& $\checkmark$ & $\checkmark$ &  &
 \\
$\mathcal{O}_{dQQeHH}$
& $\checkmark$ & $\checkmark$ &              &
&              &              & $\checkmark$ &
\\
\hline
  \end{tblr}
\end{adjustbox}
\caption{List of two-body nucleon decay modes induced by $d=8$ operators.
For further discussion, see the text.}
\label{Tab:d8channels}
\end{table}
Since there is no operator at $d=6$ that can induce $n\to K^- \ell^+$, 
a natural question to ask is: 
What is the lowest dimension required to induce such a $(B-L)$-conserving 
nucleon decay? 

At $d=8$ SMEFT has 35 operator structures with a total of 7836 real
parameters \cite{Fonseca:2017lem} that can generate nucleon decays.
We have checked that none of these operators can generate the decay
$n\to K^- \ell^+$. This check is done in the following way. 
We can divide the 35 operator structures into four sub-classes: 
$D^2\psi^4$ (4 operator structures), $X\psi^4$ (17), 
$H^2\psi^4$ (7) and $DH\psi^4$ (7). 
Here, $D$, $X$, $H$ and $\psi$ stand symbolically for a covariant
derivative, a field strength tensor, the SM Higgs field and SM
fermions. 
All terms of types $D^2\psi^4$ and $X\psi^4$ are of the form
${\cal O}_6\times D^2$ or ${\cal O}_6\times X$, where ${\cal O}_6$ is
one of the $d=6$ operators. Thus, trivially, none of them can lead
directly\footnote{This statement is strictly true only in SMEFT, see
the discussion below.} to any new two-body decay modes beyond those
shown in Tab.~\ref{Tab:d6d7channels}. 
The structures $H^2\psi^4$ can be further divided into 
$(H^{\dagger}H)\psi^4$ (4 operators) and $(HH)\psi^4$ 
or $(H^{\dagger}H^{\dagger})\psi^4$ (together 3 operator structures).  
The operators in the category of $(H^{\dagger}H)\psi^4$ do not lead to any new two-body
decays either, for the same reason as for $D^2\psi^4$ above.  
The remaining 10 operator structures, $(HH)\psi^{4}$, $(H^{\dagger}H^{\dagger})\psi^{4}$
and $DH\psi^{4}$, are listed with the two-body decay channels
that they can generate in Tab.~\ref{Tab:d8channels}.
In this table,
the operator names define the particle content in the operators, but
we do not spell out all Lorentz- and $SU(2)_{L}$-contractions, for
simplicity. Explicit operator lists can be found in
\cite{Murphy:2020rsh,Li:2020gnx,Heeck:2026ked}. 
Only two operators,
$\mathcal{O}_{DuddLH}$ and $\mathcal{O}_{ddQLHH}$, lead to
combinations of final states different from the possibilities listed
in Tab.~\ref{Tab:d6d7channels}. 
All other operators will lead to the final states $\pi^0 \ell^+$ where $\ell \in\{e,\mu\}$, 
which have the most stringent half-life limits. 
Important for the current paper, however, is that
none of these operators generates $n\to K^- \ell^+$ either.

\begin{figure}[t]
 \begin{center}
  \unitlength=1cm
  \begin{picture}(7,4)
  \put(0,0){\includegraphics[width=7cm]{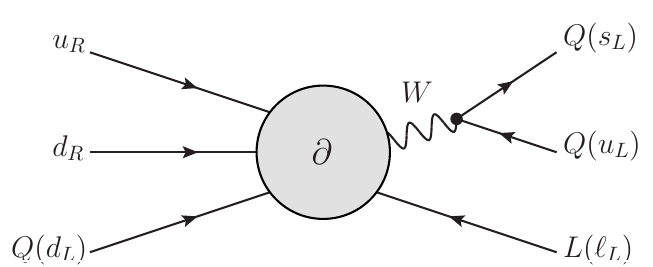}}
  \end{picture}
 \end{center}
\caption{$d=10$ operator $\partial uudd\bar{s} \ell$ generated from
  the $d=8$ operator $\mathcal{O}_{LQudW}^{(1)}$ and a SM gauge
  interaction.}
\label{Fig:d10ind8SMEFT}
\end{figure}

We stated above that $d=8$ SMEFT operators do not {\em directly}
generate new two-body final states. There is, however, a subtlety 
in this statement that should be made explicit. 
Consider, for example, the $d=8$ SMEFT operator $ \mathcal{O}_{LQudD^2}^{(1)}$
listed in \cite{Murphy:2020rsh}, which is 
\begin{equation}
\label{eq:OLQudD^21}
  \mathcal{O}_{LQudD^2}^{(1)} 
   = 
   \frac{c_{d=8}}{\Lambda^4}
   \epsilon^{IJK} \epsilon^{ij}
   D_{\rho}
   \left(
    \overline{{d_{R}}^{c}}_{I} u_{R J}
	  \right)
   D^{\rho}
   \left(
    \overline{L^{c}}_{i} Q_{K j} 
    \right)
\end{equation}
After electroweak symmetry breaking, this operator
is matched onto an effective operator in LEFT, which is proportional to 
$\partial uudd\bar{s} \ell$, as illustrated in Fig.~\ref{Fig:d10ind8SMEFT}, 
inducing then the decay $n\to K^- \ell^+$. 
Although this is a $d=10$ operator in LEFT, it is less suppressed than a standard $d=10$
SMEFT one; i.e., it scales as $G_F/\Lambda^4$ rather than as $1/\Lambda^6$.
However, the decay rate of $n\to K^- \ell^+$ generated by this
operator (and all others at $d=8$) is completely negligible, since the
operator also generates $K^0 \ell^+$ (and $\pi^0 \ell^+$) final states. A
trivial estimate by dimensional analysis leads to a ratio  of the rate
of $n\to K^- \ell^+$ relative to $p\to$ $K^0 \ell^+$ of order 
$m_p^4 G_F^2 \sim 10^{-10}$. 
Thus, $d=8$ contribution to $n\to K^- \ell^+$ is far beyond experimental reach.

\begin{figure}[t]
 \begin{center}
  \unitlength=1cm
 \begin{picture}(6,3)
  \put(0,0){\includegraphics[width=6cm]{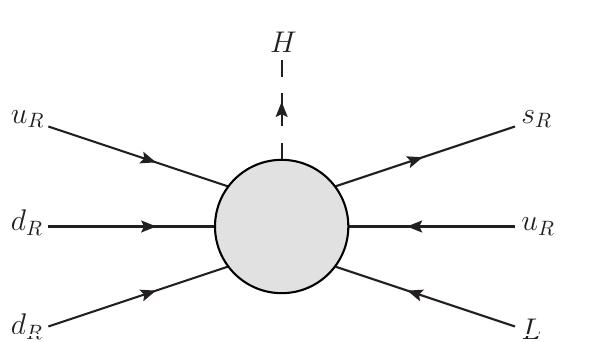}}
  \end{picture}
  \hspace{1cm}
 \begin{picture}(6,3)
  \put(0,0){\includegraphics[width=6cm]{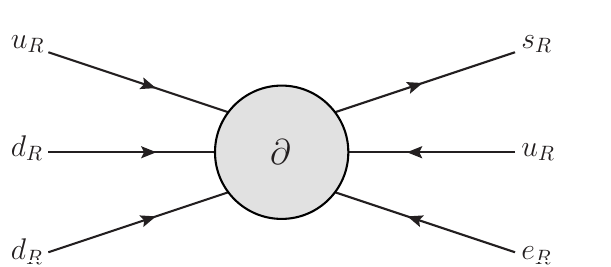}}
 \end{picture}
 \end{center}
\caption{Two simple examples of $d=10$ operators that can generate the
  $B-L$-conserving decay process $n \rightarrow K^{-}\ell^{+}$. The
  operator on the right contains a derivative.}
\label{Fig:d10ops}
\end{figure}

At $d=10$ there are already 273 operator structures which
violate $(B+L)$. Here, one encounters for the first time operators
that can generate the two-body decay $n \rightarrow K^{-} \ell^{+}$.
Two simple examples are shown in Fig.~\ref{Fig:d10ops}.
The decay rate of $n \rightarrow K^{-} \ell^{+}$ induced by the
operator with a Higgs scalar, Fig.~\ref{Fig:d10ops} left, can be
estimated with the PCAC relation as
\begin{align}
 \Gamma(n \rightarrow K^{-}\ell^{+})
 =&
 \frac{|c_{d=10}^{H}|^{2} \langle H^{0} \rangle^{2} \beta^{2}}{128 \pi \Lambda^{12}}
 \left(
 \frac{f_{K^{\pm}} m_{K^{\pm}}^{2}}{m_{u} + m_{s}}
 \right)^{2}
 m_{n}
 \left(
 1 - \frac{m_{K^{\pm}}^{2}}{m_{n}^{2}}
 \right)^{2}
 \nonumber
 \\
 =&
 \frac{1}{3.2 \cdot 10^{31}\text{yr}}
 \left(\frac{c_{d=10}^{H}}{1.0}\right)^{2}
 \left(
 \frac{1.1\cdot 10^{5}\text{GeV}}{\Lambda}
 \right)^{12}
 \label{eq:decayrate-d10H}
\end{align}
in the limit of a massless $\ell$. Here, we use the value for the
relevant hadron matrix element $\beta=0.0144$ GeV$^{3}$, which is
taken from \cite{Aoki:2017puj}.  
For the decay rate induced by the operator containing a derivative instead of the Higgs, 
one should replace $\langle H^{0} \rangle$ by the momentum of one of the final state
particles. 
In the two-body decay $n\to K^{-} \ell^{+}$, 
the momenta are roughly $p_{K}=p_{\ell} \simeq 0.3 m_n$, 
leading to a correspondingly smaller rate for this operator. 
Note that both operators also generate  the decay $p\to K^0 \ell^{+}$ with the same rate.

Assuming a common Wilson coefficient, the estimated scale $\Lambda$ 
for a $d=10$ operator is more than five orders of magnitude lower than 
the one inferred for the $d=7$ induced decay $n\to K^+ \ell^-$, 
see Eq.~\eqref{eq:DecayRate-d7-n-to-Kl}.
From an EFT point of view  a $d=7$ operator would therefore appear to be 
the natural explanation for a hypothetical  observation of $n\to K \ell$. 
However, $\Lambda \sim 100$ TeV
is far above the energy scales that can be probed in direct experiments,
such as the LHC. Thus, a priori, it seems impossible to
exclude $d=10$ operators as the origin of $n\to K \ell$.

\begin{figure}[t]
 \begin{center}
  \unitlength=1cm
 \begin{picture}(7,6)
  \put(0,0){\includegraphics[width=7cm]{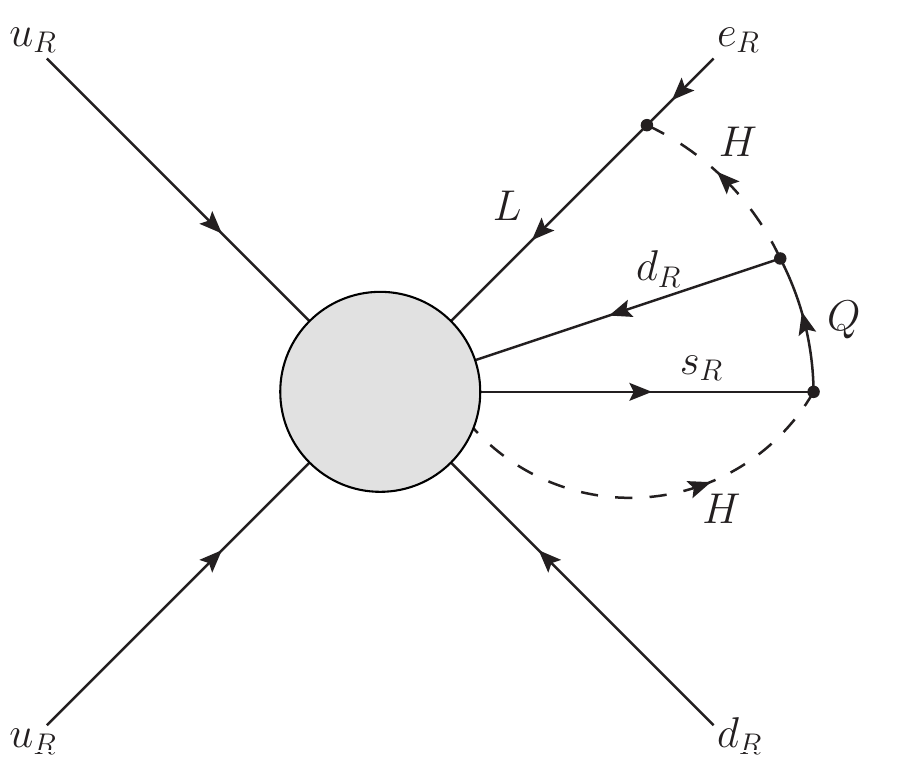}}
 \end{picture}
  \hspace{0.5cm}
  \begin{picture}(7,6)
  \put(0,0){\includegraphics[width=7cm]{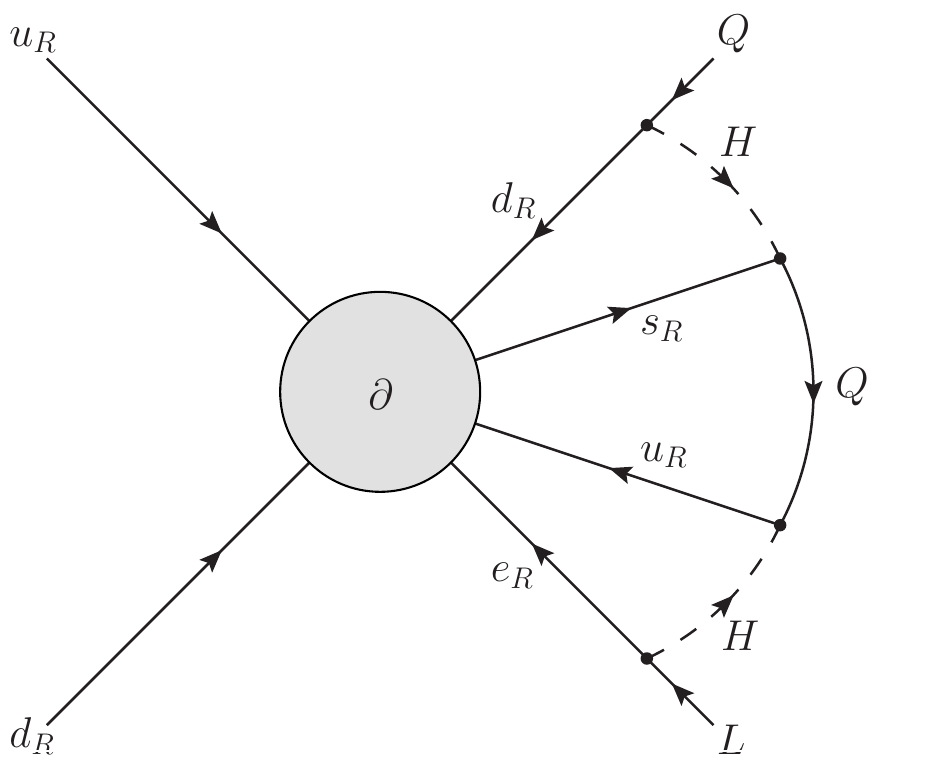}}
 \end{picture}
 \end{center}
\caption{Examples of ``black-box'' diagrams for the $d=10$ operators,
  shown in Fig.~\ref{Fig:d10ops}.  The 3-loop diagrams result in the
  generation of $d=6$ operators.}
\label{Fig:d10BB}
\end{figure}
The EFT argument in favour of a $d=7$ origin can, however, be strengthened as follows.
All $d=10$ operators violate $(B+L)$ by two units, while conserving
$(B-L)$. Therefore, if a $d=10$ operator is allowed by the symmetries
of the underlying Lagrangian, one expects that also the $d=6$
operators (obeying also $(B+L)=2$ and $(B-L)=0$) are allowed by the
same symmetries. 
These $d=6$ operators would then dominate the decay rate by a large factor. 
Thus, forbidding the $d=6$ operators while allowing the $d=10$ ones, 
 would require an additional symmetry responsible for the absence of 
the $d=6$ operators. We will now show, in a ``proof by contradiction'', 
that no such symmetry can exist.  
This chain of reasoning follows the same logic as the well-known black-box 
theorem of neutrinoless double beta decay~\cite{Schechter:1981bd,Hirsch:2006yk}.
Let us assign charges $q_{i}$ to each SM field $i\in\{L, Q, e_{R},
u_{R}, d_{R}, H\}$. The presence of the $d=10$ operator in the left panel
of Fig.~\ref{Fig:d10ops} imposes the following constraint:
\begin{equation}\label{eq:d10C}
2 q_{u_R} + 2 q_{d_R} - q_{s_R} + q_{L} - q_H =0.
\end{equation}  
To forbid the $d=6$ operator ${\cal O}_{duue}$, on the other hand,
one must require
\begin{equation}\label{eq:d6C}
2 q_{u_R} + q_{d_R} + q_{e_R}  \ne 0.
\end{equation}  
The SM Yukawa interaction $\overline{L}e_RH$ implies
$q_{e_R}=q_L-q_H$. Therefore, Eq.~\eqref{eq:d10C} can be written as
\begin{equation}
\left(2q_{u_R}+q_{d_R}+q_{e_R}\right)
+\left(q_{d_R}-q_{s_R}\right)=0 .
\end{equation}
Thus, if ${\cal O}_{duue}$ is to be forbidden while the $d=10$ operator is allowed, 
one must have
\begin{equation}
\label{eq:cons}
q_{d_R} - q_{s_R}  \ne 0.
\end{equation}  
However, since  the CKM matrix is non-trivial, the non-diagonal Yukawa interactions 
have to be non-zero as well, i.e., $\overline{Q_{\alpha}} d_{R \beta} H$ must be allowed. 
This, together with the usual down-quark Yukawa interaction, implies 
$q_{d_R} = q_{s_R}$ and completes the proof.

Thus far, we have shown that $d=6$ operators must co-exist with
$d=10$ operators. Nevertheless, the above discussion does not fix
the relative size of their Wilson coefficients. It could still
be true that for some unknown reason $c_{d=6}$ is many orders of
magnitude smaller than $c_{d=10}$. This would correspond to a highly
fine-tuned situation, but again, it is not excluded a priori.

Still, we can show at least that $c_{d=6}$ cannot be exactly zero 
when $d=10$ operators exist.  
All $d=10$ operators allow to close some quark (and
lepton) lines, see Fig.~\ref{Fig:d10BB}, to generate lower
dimensional operators radiatively. 
The figure shows how the two $d=10$ operators from Fig.~\ref{Fig:d10ops} generate 
the $d=6$ ${\cal O}_{duue}$ and ${\cal O}_{duQL}$ ones at the order of 3-loop.
This can be used to establish lower bounds on the corresponding $c_{d=6}$ Wilson coefficients, assuming that the associated $c_{d=10}$ are non-zero.
A complete treatment of this effect would require deriving and running 
the renormalization group equations for the relevant operators, 
a task beyond the scope of this paper. 
Instead, we perform a very rough order-of-magnitude estimate of the 3-loop diagrams, 
obtaining
\begin{align}
\label{eq:3lp1}
c_{duue} \sim& \frac{1}{(16 \pi^2)^3} y_ey_dy_s c^H_{d=10}, \\ 
\label{eq:3lp2}
c_{duQL} \sim& \frac{1}{(16 \pi^2)^3} y_ey_dy_sy_u c^{\partial}_{d=10}.
\end{align}
While the numerical values implied are very small numbers, 
the limit on the $d=6$ operators is also very strong, 
compare to the limits shown in Tab.~\ref{Tab:d6d7channels}.
Setting $c^H_{d=10}=1$ in Eq.~\eqref{eq:decayrate-d10H}, one finds that a scale
$\Lambda$ of order $100~{\rm TeV}$ gives a half-life for $n\to K^-\ell^+$ comparable 
to the current bound,
$\tau(n\to K^-\ell^+)\sim 3.2\times 10^{31}~{\rm ys}$.
Using then $c^H_{d=10}=1$ in
Eq.~\eqref{eq:3lp1}, 
together with corresponding values of the Yukawa couplings, 
gives the three-loop estimate $c_{d=6}^{\text{3-loop}}\simeq 10^{-20}$. 
However, the upper limit on $c_{ddue}$ for $\Lambda=100$ TeV from the
half-life limit of $p \to \pi^0 e^+ \sim 2.4\times 10^{34}$ ys turns out 
to be $c_{ddue} \le 4.5 \times 10^{-22}$, 
which is around a factor of $\sim 20$ smaller. 
This would exclude ${\cal O}_{d=10}^H$ as a possible explanation for $n\to K \ell$. This conclusion, however, is only valid for $c_{d=10}^{H} \gtrsim 10^{-2}$. For smaller values, the induced three-loop coefficient $c_{6}^{3{\rm -loop}}$ falls below the upper limit on $c_{ddue}$.
For ${\cal O}_{d=10}^{\partial}$, however, the situation is not that
straight-forward, since $c_{d=6}^{\text{3-loop}} \simeq 10^{-25}$ for
$c^{\partial}_{d=10}=1$. 
This is too small compared to the limit on $c_{duQL}$ for $\Lambda=100$ TeV 
and thus ${\cal O}_{d=10}^{\partial}$
cannot be excluded as the origin of $n\to K \ell$ by this argument.

In summary, if a ${\cal O}_{d=10}$ operator is used to generate
a non-zero $n\to K^- \ell^+$ with an observable life-time, 
all Wilson coefficients for $d=6$ operators need to have values
smaller than, very roughly, $c_{d=6} \lsim 10^{-22}$, otherwise
$p\to \pi^0 e^+$ should already have been observed. 
We think this argument is strong enough to claim that the observation of
$n\to K \ell$ would point to $(B-L)$ violation, even if the charges
of the lepton (and kaon) are not observed experimentally.

\section{Constraints on baryogenesis} 

In this section, we discuss briefly the consecuences for high-scale
baryogenesis scenarios that could be derived from an observation of a
$(B-L)$-violating nucleon decay. We will concentrate on constraints on
$(B-L)$-violating operators, that follow from the requirement that the
operator does not lead to a washout of a pre-existing baryon
asymmetry, see for example ~\cite{Campbell:1990fa,Campbell:1991at,
  Fukugita:1990gb, Fischler:1990gn,Deppisch:2013jxa,Deppisch:2015yqa}.
This constraint is relevant for a large class of baryogenesis
scenarios based on baryon number generation at high temperatures, such
as GUT baryogenesis
~\cite{Yoshimura:1978ex,Weinberg:1979bt,Kolb:1983ni} or standard
thermal leptogenesis ~\cite{Fukugita:1986hr,Flanz:1994yx, Covi:1996wh,Pilaftsis:1997jf,Buchmuller:1997yu,Giudice:2003jh}, among others.  We will not
attempt to construct a successful baryogenesis model from
$(B-L)$-violating operators, for earlier work along these lines, see
for example ~\cite{Enomoto:2011py,Babu:2012vb,Babu:2012iv,Hati:2018cqp}.

For concreteness, we will concentrate on a particular operator,
$\mathcal{O}_{\overline{L}dddH}$, in the following. However, the
arguments presented below are valid for any of the $d=7$ operators
listed in Tab.~\ref{Tab:d6d7channels}, up to some numerical
factors. Consider the decomposition of $\mathcal{O}_{\overline{L}dddH}$.
We do not exhaustively list all possible tree-level decompositions
of this $d=7$ operator, since they can be found in \cite{Li:2023cwy}. 
However, at tree-level there are just two types of diagrams, see
Fig.~\ref{Fig:d7decom-tree}: Class-I contains a heavy scalar and a
BSM  fermion, while Class-II contains two BSM scalars.
The necessary interactions for those example diagrams are:
\begin{figure}[t]
\begin{center}
 \unitlength=1cm
\begin{picture}(6,4)
 \put(0,0){\includegraphics[width=6cm]{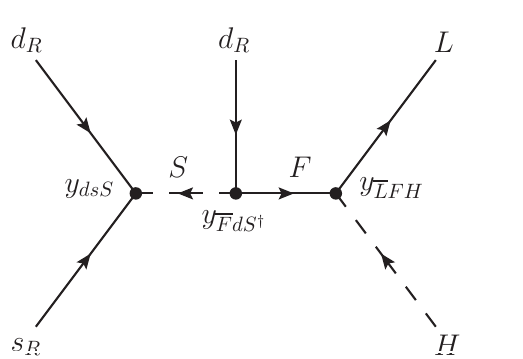}}
 \put(2.1,0){Class-I}
\end{picture}
\begin{picture}(6,4)
 \put(0,0){\includegraphics[width=6cm]{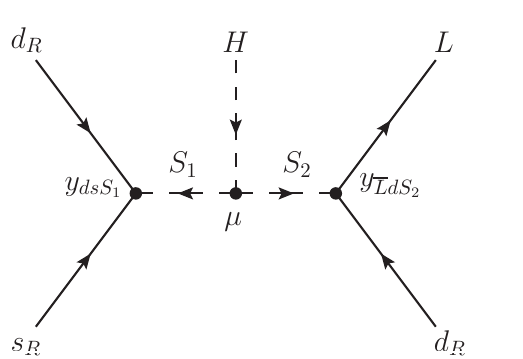}}
 \put(2.1,0){Class-II}
\end{picture}
\end{center}
\caption{Example diagrams for the two possible tree-level decompositions
  of the effective operator $\mathcal{O}_{\overline{L}dddH}$.}
\label{Fig:d7decom-tree}
\end{figure}
%
\begin{align}
 \mathscr{L}_{\text{I}}
 =&
 y_{dsS}
 \epsilon^{IJK}
 (\overline{{d_{R}}^{c}})_{I \dot{a}}
 (s_{R})_{J}^{\dot{a}}
 S_{K}
 +
 y_{\overline{L}FH}
 (\overline{L})^{i}_{\dot{a}}
 (F_{R})^{\dot{a}}
 H_{i}
 +
 y_{\overline{F}d S^{\dagger}}
 (\overline{F_{L}})_{\dot{a}}
 (d_{R})_{I}^{\dot{a}}
 S^{\dagger I},
\label{eq:d7-decom-tree-example-Nr1}
\\
 \mathscr{L}_{\text{II}}
 =&
 y_{dsS_{1}}
 \epsilon^{IJK}
 (\overline{{d_{R}}^{c}})_{I \dot{a}}
 (s_{R})_{J}^{\dot{a}}
 (S_{1})_{K}
 +
 y_{\overline{L}dS_{2}}
 (\overline{L})^{i}_{\dot{a}}
 (d_{R})_{I}^{\dot{a}}
 (S_{2})^{I}_{i}
 +
 \mu
 (S_{1}^{\dagger})^{I}
 (S_{2}^{\dagger})_{I}^{i}
 H_{i}.
\label{eq:d7-decom-tree-example-Nr2}
\end{align}
Here , $S(=S_{1})$ is a scalar with SM charges $(\vec{3},\vec{1},+2/3)$,
$F$ is a vector-like fermion with $(\vec{1},\vec{1},-1)$,
and  $S_{2}$ is a scalar with $(\overline{\vec{3}},\vec{2},-1/6)$.
%
The $d=7$ operator can be matched to this Lagrangian as
\begin{equation}\label{eq:mtch}
\frac{c_{\overline{L}dddH}}{\Lambda^3} = \tilde{c}_{d=7} \frac{\tilde{\mu}}{\Lambda^4},
\end{equation}
where in the effective field theory limit we assume, as usual,
$m_F = m_{S_i} = \Lambda$ and 
\begin{equation}\label{eq:mtch2}
\begin{array}{lll}
 \tilde{c}_{d=7} = y_{dsS} \cdot y_{\overline{F}dS^{\dagger}} \cdot y_{\overline{L}FH}, 
 & \quad\tilde{\mu} = m_F,
  &\quad\text{for Class-I},
 \\
 \tilde{c}_{d=7} = y_{dsS_{1}} \cdot y_{\overline{L}dS_{2}},  
  &\quad\tilde{\mu} = \mu,  
  &\quad\text{for Class-II}.
\end{array}
\end{equation}
Note that the parameter $\mu$ in class-II could, in principle, be
much smaller than $\Lambda$. 

Different processes can lead to changes in baryon and lepton
number. The most relevant channel for the washout of $(B-L)$ is the
$s$-channel scattering process $d s \leftrightarrow \bar{d} \ell
H^{0}$, which is resonantly enhanced at temperatures close to the mass
of the mediator fields. For mediator masses all equal to $\Lambda$,
the cross section of the $s$-channel process in both cases of
Eq.~\eqref{eq:d7-decom-tree-example-Nr1} and
Eq.~\eqref{eq:d7-decom-tree-example-Nr2} can be written
as\footnote{This cross section formula diverges at the temperature
that satisfies $\sqrt{s}=\Lambda$. As usual, in s-channel processes
this divergence has to be regulated with the width of the mediator
fields.  We are mostly interested in the edge of the parameter regions
where the rate $\Gamma$ becomes larger than the Hubble parameter.  For
that purpose, including the width of the mediator fields is not
important.}
\begin{align}
\sigma
=
\frac{1}{6144\pi^{3}}
\frac{\left|\tilde{\mu} \tilde{c}_{d=7}\right|^{2}}{\left(s-\Lambda^{2}\right)^{4}}
s^{2},
\label{eq:crosssection-2-to-3}
\end{align} 
where $\sqrt{s}$ is the centre-of-mass energy of the initial fermions.
The average energy $\langle E \rangle$ of the fermions in the thermal bath
at a temperature $T$ can be estimated by the energy density divided by
the number density of the Fermi-Dirac distribution, which is $\langle
E \rangle \simeq 3.15 T$, and $\sqrt{s} = 2 \langle E \rangle$.  The
coefficient $\tilde{c}_{d=7}$ and the mass parameter $\tilde{\mu}$ are
given in Eq.~\eqref{eq:mtch2}.  Note that the coefficient $c_{d=7}$ in
the decay rate formula Eq.~\eqref{eq:DecayRate-d7-n-to-Kl} of $n
\rightarrow K^{+} \ell^{-}$ is related to $\tilde{c}_{d=7}$ via
$c_{d=7} = \tilde{\mu}\tilde{c}_{d=7}/\Lambda$.

The reaction rate $\Gamma$ of the 2-to-3 scattering process can be
estimated from the product of the number density of the initial
fermions $n = 2 \cdot \frac{3\zeta(3)}{4 \pi^{2}} T^{3}$ and the cross
section $\sigma$ in Eq.~\eqref{eq:crosssection-2-to-3}.  Comparing
$\Gamma=n\cdot \sigma$ with the Hubble parameter $H = 1.66
\sqrt{g_{*}} \frac{T^{2}}{M_{\text{Pl}}}$, one can estimate whether
the process is in thermal equilibrium. Here, $g_{*}$ denotes the
number of effective degrees of freedom, with $g_{*}\simeq 106.75$ when
counting only SM particles, and $M_{\text{Pl}} =1.22 \cdot 10^{19}$
GeV is the Planck mass.

Note that in the limit where $T \ll \Lambda$, $\Gamma/H$ scales
proportional to $\Gamma/H \propto T^5$, while in the limit $\Lambda
\ll T$ one obtains $\Gamma/H \propto T^{-3}$. This can be understood
as follows. In the high temperature regime, the universe expands too
rapidly and the reaction $d s \leftrightarrow \bar{d} L H^{\dagger}$
is too slow to come into equilibrium, even for practically massless
mediators. Once the universe has cooled down (and slowed down its
expansion, $H \propto T^2$) the process enters an equlibrium phase.
On the other hand, in the low temperature regime the cross section is
suppressed by the scale $\Lambda$ and thus there is a critical
temperature, below which the reaction $d s \leftrightarrow \bar{d} L
H^{\dagger}$ drops out of equilibrium again. This defines a
temperature range, in which any pre-existing baryon asymmetry is
washed out. This range depends on the choice for $\tilde{\mu}$ and
$\tilde{c}_{d=7}$.

\begin{figure}[t]
\begin{center}
\unitlength=1cm
\begin{picture}(6.5,7)
 \put(0,0){\includegraphics[width=6.5cm]{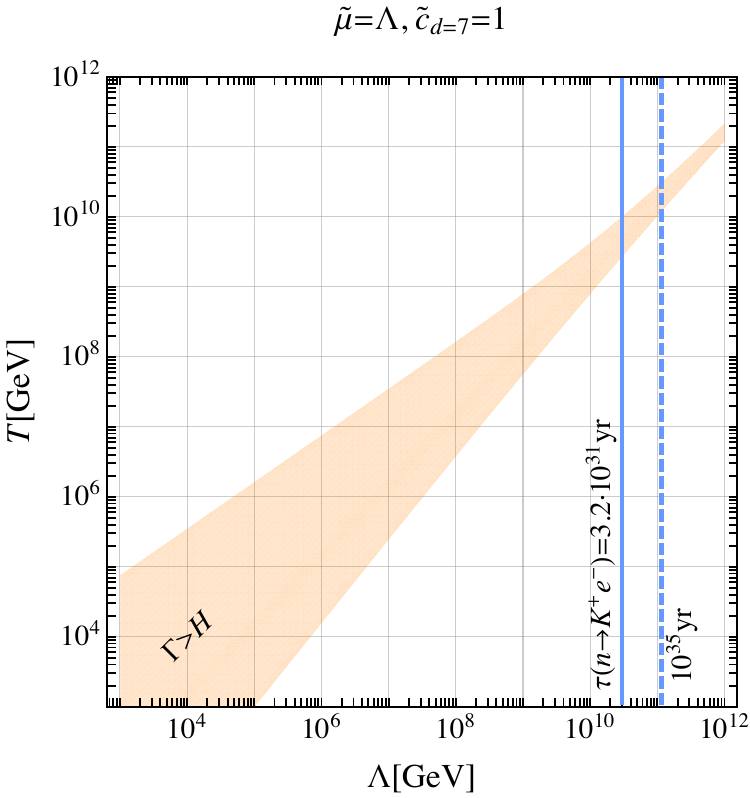}}
\end{picture}
\hspace{0.5cm}
\begin{picture}(6.5,7)
 \put(0,0){\includegraphics[width=6.5cm]{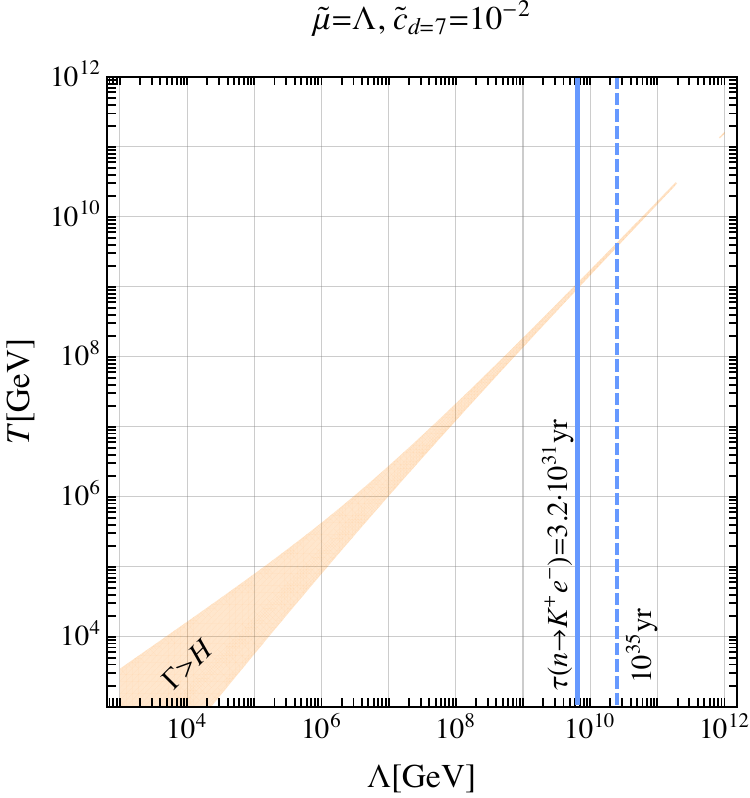}}
\end{picture}
\end{center}
\caption{Contours of $\Gamma=H$ in the plane $\Lambda$-$T$ for two
  different choices of $\tilde{\mu}$ and $\tilde{c}_{d=7}$. Within the
  shaded regions $\Gamma>H$ and thus, the $(B-L)$-violating process $d s
  \leftrightarrow \bar{d} L H^{\dagger}$ is in thermal equilibrium.
  The two vertical lines correspond to the life-times of $\tau(n\to
  K^+ e^-)$ of $3.2\times 10^{31}$ and $10^{35}$ yr.}
\label{Fig:Gamow-Lambda-vs-T}
\end{figure}

In Fig.~\ref{Fig:Gamow-Lambda-vs-T} we plot contours of $\Gamma = H$
in the plane $\Lambda$-$T$ for two different choices for $\tilde{\mu}$
and $\tilde{c}_{d=7}$. In the orange shaded region, the reaction rate
$\Gamma$ becomes larger than the Hubble expansion rate $H$,
i.e. within this parameter region the $(B-L)$-violating process $d s
\leftrightarrow \bar{d} L H^{\dagger}$ is in equilibrium and leads to
efficient wash-out.  The vertical lines indicate the scales $\Lambda$
that correspond to two fixed life-times for the decay $n \rightarrow
K^{+} \ell^{-}$.  The current bound ($3.2\cdot 10^{31}$ yr) is
indicated with a solid line, and we also show with a dashed line
$\Lambda$ for $10^{35}$ yr, as a reference value for some future
sensitivity.

Note that the cross section, see Eq.~\eqref{eq:crosssection-2-to-3},
is proportional to $\sigma \propto \left|\tilde{\mu}
\tilde{c}_{d=7}\right|^{2}$. Thus, choosing for example
$\tilde{\mu}=10^{-2}\Lambda$ and $\tilde{c}_{d=7}=1$ leads to the same
plot as in Fig.~\ref{Fig:Gamow-Lambda-vs-T}, to the right. From this
plot one can also see that for values of $|\tilde{c}_{d=7}|$ smaller
than, very roughly, (few) $10^{-3}$, the region where $\Gamma>H$
occurs is confined to values of $\Lambda$ already rule out by the
current bound on $\tau(n\to K^+ e^-)$. The requirement that a
baryon number generated at very high temperatures survives to the
present day, can be interpreted as an upper bound on $|\tilde{c}_{d=7}|$,
if $n\to K^+ e^-$ is discovered in the future.

On the other hand, if $|\tilde{c}_{d=7}\tilde{\mu}|$ is larger than
(roughly) $10^{-2}\Lambda$ and the universe went through a phase with
a reheating temperature larger than the orange regions in
Fig.~\ref{Fig:Gamow-Lambda-vs-T}, any pre-existing baryon asymmetry is
washed out. Thus, in this case, the baryon number observed today must
be generated below that critical temperature, i.e.  baryogenesis
should occur below temperatures in the range $10^{9}$-$10^{10}$ GeV,
depending on the actual lifetime for $n\to K^+ e^-$.  This would
disfavour GUT baryogenesis, or also standard thermal leptogenesis.

\begin{figure}[t]
 \unitlength=1cm
\begin{center}
 \begin{picture}(8,8)
  \put(0,0){\includegraphics[width=8cm]{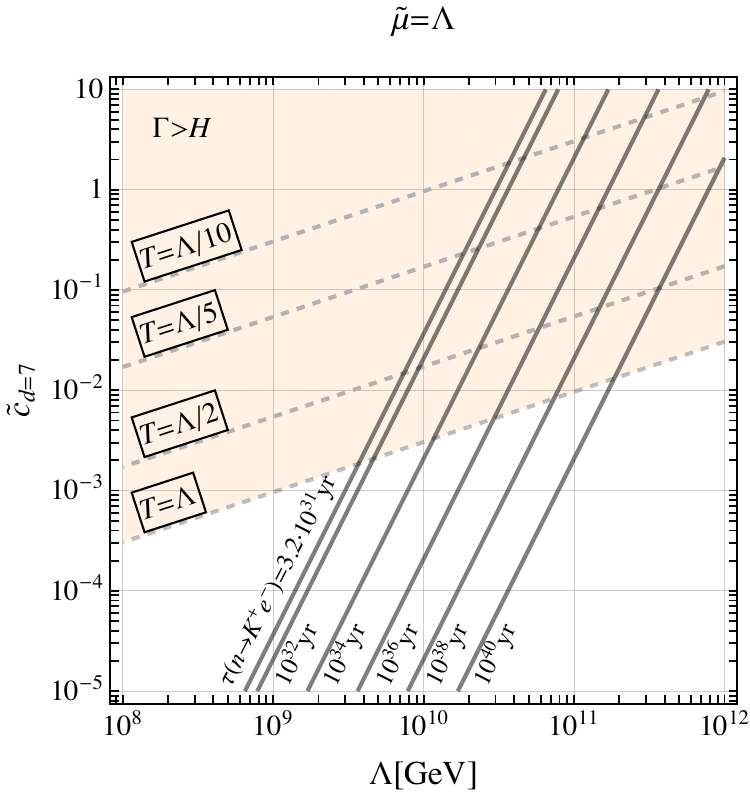}}
 \end{picture}
\end{center}
\caption{Comparison between the contours of $\Gamma=H$ and the
  lifetime of $n \rightarrow K^{+} e^{-}$, with the assumption of
  $\tilde{\mu} = \Lambda$.}
\label{Fig:Lambda-c7}
\end{figure}
To illustrate better the parameter regions explored by nucleon decay
searches and baryogenesis, we plot $\Gamma/H$ and the lifetime of $n
\rightarrow K^{+} e^{-}$ in the plane $\Lambda$-$\tilde{c}_{d=7}$ in
Fig.~\ref{Fig:Lambda-c7}.  In this plot, we fix $\tilde{\mu} =
\Lambda$.  Here, we use the cross section formula in the
low-temperature limit $\sigma \simeq \frac{1}{6144 \pi^{3}}
\frac{|\tilde{\mu} \tilde{c}_{d=7}|^{2}}{\Lambda^{8}} s^{2}$ which is
valid for $T \lesssim \Lambda$ and set $T=f \cdot \Lambda$, with
$f=(1/10,1/5,1/2,1)$. The parameter combinations that lead to
$\Gamma/H=1$ are marked with $T=f \cdot \Lambda$ and shown as dashed
gray lines. For values of $\tilde{c}_{d=7}$ larger than the dashed
lines $\Gamma/H > 1$ and thus, this part of parameter space is
constrained by the washout argument, if the reheating temperature is
larger than the value indicated. As explained above, the strongest
bound on $\tilde{c}_{d=7}$ is obtained for $T=\Lambda$.

Once $n \rightarrow K^{+} e^{-}$ is discovered with a certain half-live,
$c_{d=7}/\Lambda^3$ is fixed. This constraint is shown for various
example lifetimes as full gray lines in Fig.~\ref{Fig:Lambda-c7}. 
The crossing point with the equilibrium condition $\Gamma = H$ 
shows the critical value of ${\tilde c}_{d=7}$, above which any
pre-existing baryon asymmetry is washed out. This upper limit
depends on the value of the reheating temperature of the universe,
as indicated by the dashed lines. If the reheating temperature
is much larger than $\Lambda$, the choice of $\Lambda=T$ for
deriving the constraint becomes conservative. If the reheat temperature never reached $T=\Lambda$, the bounds on ${\tilde c}_{d=7}$ will be weaker, as indicated in the figure.

To summarize, a discovery of $n \rightarrow K \ell$ would provide
interesting information for models of baryogenesis. As we discussed,
an upper limit on the reheat temperature of the universe can be
derived, from the requirement the $(B-L)$ violation does not come into
equilibrium. High-scale models of baryogenesis can be constrained from
this argument, if the relevant $d=7$ operators have Wilson
coefficients larger than roughly $c_{d=7} \sim 10^{-2}$.

\section{Summary}
\label{Sec:summary}

In this paper we have studied neutron decay into a charged kaon and a
charged lepton, with emphasis on what such a signal would imply for the origin of the baryon asymmetry of the
universe. The main observation is that the final state $n\to K\ell$
contains more information than may be apparent experimentally. If the
charge of the kaon or of the charged lepton is not measured, the two
possibilities $n\to K^+\ell^-$ and $n\to K^-\ell^+$ may look identical. In SMEFT, however, they have quite different interpretations.
The decay $n\to K^+\ell^-$ violates $(B-L)$ and can be generated by
dimension-seven operators, while the $(B-L)$-conserving mode
$n\to K^-\ell^+$ first arises, as a genuine two-body decay, from
operators of dimension ten.
This difference in operator dimension is numerically important. For Wilson coefficients
of comparable size, the dimension-seven contribution probes scales more than five
orders of magnitude higher than the scale associated with the dimension-ten operator. Therefore, an
observation of $n\to K\ell$ would most naturally point to a dimension-seven,
and hence $(B-L)$-violating, origin. A dimension-ten, $(B-L)$-conserving
interpretation would be possible only in non-generic scenarios in which
the accompanying lower-dimensional effects are sufficiently suppressed. We therefore conclude that an observation of $n\to K\ell$, especially
in the absence of accompanying modes such as $p\to \pi^0\ell^+$, would
strongly suggest that $(B-L)$ is violated. This statement is not a
model-independent theorem, since dimension-ten operators at scales of
order $100$ TeV can still be viable in highly fine-tuned arranged scenarios.
Nevertheless, the dimension-seven interpretation appears to be the more
natural EFT explanation.

We have also discussed the implications for baryogenesis. If the same
$(B-L)$-violating interaction responsible for $n\to K^+\ell^-$ was in
thermal equilibrium before the electroweak phase transition, it would
erase any pre-existing $(B-L)$ asymmetry. Together with sphaleron
processes, this would wash out a baryon asymmetry generated at higher
temperatures. A future discovery of $n\to K\ell$ would therefore
constrain the thermal history of the universe. The strength of this
constraint depends on the relevant Wilson coefficient and on the
maximum temperature reached after inflation. Certain choices
of parameters can lead to efficient
washout in the region probed by nucleon decay searches. In that part of
parameter space, high-scale baryogenesis scenarios, such as GUT
baryogenesis or standard thermal leptogenesis, would not work if the reheat temperature does not reach $T = m_{GUT}$.

In summary, $n\to K\ell$ is a particularly interesting channel
to search for. The existing limits are old and considerably weaker than
those on many standard nucleon decay modes. A dedicated analysis by
Super-Kamiokande, and later by Hyper-Kamiokande, DUNE or JUNO, could
improve the sensitivity substantially. Such a search would test baryon
number violation in a channel that is especially sensitive to
$(B-L)$-violating physics and, indirectly, to the possible temperature
scale of baryogenesis.

\bigskip
\centerline{\bf Acknowledgements}

\medskip

M.H thanks Renato Fonseca for many discussions on group theory and
help with \texttt{Sym2Int} \cite{Fonseca:2017lem}.
J.C.H and T.O acknowledge support from ANID – Millennium Science
Initiative Program ICN2019 044.  
The research of J.C.H is supported by
ANID Chile through FONDECYT regular grant N${}^{\underline{\text{o}}}$
1241685.
The research of T.O. is supported by
ANID Chile through FONDECYT regular grant N${}^{\underline{\text{o}}}$
1250343. 
M.H. acknowledges support by Spanish grants
PID2023-147306NB-I00 and CEX2023-001292-S
(MCIU/AEI/10.13039/501100011033).
C.A. is supported by 
ANID Chile through FONDECYT regular grant N${}^{\underline{\text{o}}}$ 1231248 and ANID CCTVal CIA250027.


\appendix
\bigskip

\bibliography{NtoKlforBAU}
\bibliographystyle{JHEP}

\end{document}